\documentclass[12pt]{article}
\usepackage{epsfig}
\usepackage{cite}

\begin{document}
\baselineskip 0.7cm

\newcommand{\gsim}{ \mathop{}_{\textstyle \sim}^{\textstyle >} }
\newcommand{\lsim}{ \mathop{}_{\textstyle \sim}^{\textstyle <} }
\newcommand{\vev}[1]{ \left\langle {#1} \right\rangle }
\newcommand{\lsp}{ \left ( }
\newcommand{\rsp}{ \right ) }
\newcommand{\lmp}{ \left \{ }
\newcommand{\rmp}{ \right \} }
\newcommand{\llp}{ \left [ }
\newcommand{\rlp}{ \right ] }
\newcommand{\labs}{ \left | }
\newcommand{\rabs}{ \right | }
\newcommand{\EV} { {\rm eV} }
\newcommand{\KEV}{ {\rm keV} }
\newcommand{\MEV}{ {\rm MeV} }
\newcommand{\GEV}{ {\rm GeV} }
\newcommand{\TEV}{ {\rm TeV} }
\newcommand{\YR}{ {\rm yr} }
\newcommand{\mgut}{M_{GUT}}
\newcommand{\mint}{M_{I}}
\newcommand{\mgra}{M_{3/2}}
\newcommand{\mll}{m_{\tilde{l}L}^{2}}
\newcommand{\mdr}{m_{\tilde{d}R}^{2}}
\newcommand{\mllXX}[1]{m_{\tilde{l}L , {#1}}^{2}}
\newcommand{\mdrXX}[1]{m_{\tilde{d}R , {#1}}^{2}}
\newcommand{\mgy}{m_{G1}}
\newcommand{\mgl}{m_{G2}}
\newcommand{\mgc}{m_{G3}}
\newcommand{\nuR}{\nu_{R}}
\newcommand{\slL}{\tilde{l}_{L}}
\newcommand{\slLi}{\tilde{l}_{Li}}
\newcommand{\sdR}{\tilde{d}_{R}}
\newcommand{\sdRi}{\tilde{d}_{Ri}}
\newcommand{\e}{{\rm e}}
\newcommand{\bsub}{\begin{subequations}}
\newcommand{\esub}{\end{subequations}}
\newcommand{\wt}{\widetilde}
\newcommand{\tm}{\times}
\newcommand{\ra}{\rightarrow}
\newcommand{\del}{\partial}
\newcommand{\az}{a_{Z}^{}}
\newcommand{\bz}{b_{Z}^{}}
\newcommand{\cz}{c_{Z}^{}}
\newcommand{\aw}{a_{W}^{}}
\newcommand{\bw}{b_{W}^{}}
\newcommand{\dw}{d_{W}^{}}
\newcommand{\sw}{s_{W}}
\newcommand{\cw}{c_{W}}
\newcommand{\gz}{g_{Z}^{}}
\newcommand{\mz}{m_{Z}^{}}
\newcommand{\pH}{p_{H}^{}}
\newcommand{\pone}{p_{1}^{}}
\newcommand{\ptwo}{p_{2}^{}}
\newcommand{\pt}{\partial}
\newcommand{\btable}{\begin{table}[htbp]\begin{center}}
\newcommand{\etable}[1]{ \end{tabular}\caption{#1}\end{center}\end{table} }
\newcommand{\vt}{\vspace{3mm}}
\renewcommand{\thefootnote}{\fnsymbol{footnote}}
\setcounter{footnote}{1}

%%%----------------------------------------
\makeatletter
%%%%%%%%%%%%%%%%%%%%  subequations  %%%%%%%%%%%%%%%%%%%%
% subequations
%%% File: subeqn.sty
%%% The subequations environment %%%
%
% Within the subequations environment, the only change is that
% equations are labeled differently.  The number stays the same,
% and lower case letters are appended.  For example, if after doing
% three equations, numbered 1, 2, and 3, you start a subequations
% environment and do three more equations, they will be numbered
% 4a, 4b, and 4c.  After you end the subequations environment, the
% next equation will be numbered 5.
%
% Both text and equations can be put inside the subequations environment.
%
% If you make any improvements, I'd like to hear about them.
%
% Stephen Gildea
% MIT Earth Resources Lab
% Cambridge, Mass. 2139
% mit-erl!gildea
% gildea@erl.mit.edu
%
\newtoks\@stequation

\def\subequations{\refstepcounter{equation}%
  \edef\@savedequation{\the\c@equation}%
  \@stequation=\expandafter{\theequation}%   %only want \theequation
  \edef\@savedtheequation{\the\@stequation}% %expanded once
  \edef\oldtheequation{\theequation}%
  \setcounter{equation}{0}%
  \def\theequation{\oldtheequation\alph{equation}}}

\def\endsubequations{%
  \ifnum\c@equation < 2 \@warning{Only \the\c@equation\space subequation
    used in equation \@savedequation}\fi
  \setcounter{equation}{\@savedequation}%
  \@stequation=\expandafter{\@savedtheequation}%
  \edef\theequation{\the\@stequation}%
  \global\@ignoretrue}

%%%%%%%%%%%%%%%%%%%%%%%%%%%%%%%%%%%%%%%%%%%%%%%%%%%%%%%%%%%%%%%%%%%%%%%%%%
%  The following change of the eqnarray environment makes the spacing
%  associated with the alignment operation && narrower.

\def\eqnarray{\stepcounter{equation}\let\@currentlabel\theequation
\global\@eqnswtrue\m@th
\global\@eqcnt\z@\tabskip\@centering\let\\\@eqncr
$$\halign to\displaywidth\bgroup\@eqnsel\hskip\@centering
%  $\displaystyle\tabskip\z@{##}$&\global\@eqcnt\@ne
%  \hskip 2\arraycolsep \hfil${##}$\hfil
%  &\global\@eqcnt\tw@ \hskip 2\arraycolsep $\displaystyle\tabskip\z@{##}$\hfil
     $\displaystyle\tabskip\z@{##}$&\global\@eqcnt\@ne
      \hfil$\;{##}\;$\hfil
     &\global\@eqcnt\tw@ $\displaystyle\tabskip\z@{##}$\hfil
   \tabskip\@centering&\llap{##}\tabskip\z@\cr}

\makeatother

%%%%
%%%%

\begin{titlepage}

\begin{flushright}
UT-02-30
\end{flushright}

\vskip 0.35cm
\begin{center}
{\large \bf New Potential of Black Holes : Quest for TeV-Scale Physics
 \\ by Measuring Top Quark Sector using Black Holes}

\vskip 0.4cm

Yosuke Uehara

\vskip 0.4cm

{\it Department of Physics, University of Tokyo, 
         Tokyo 113-0033, Japan}\\
\vskip 1.5cm

\abstract{If $\TEV$-scale gravity models are correct, the production
of black holes will be the first signal of new physics. Once black holes
are produced, they will give us much information about $\TEV$-scale
new physics directly. But such black holes can also be used for the precision
measurements of the Standard Model (SM). The SM is nothing but a theory
which can describe weak-scale, and $\TEV$-scale physics will affect it.
So if some experimental results which cannot be explained in the SM are
found, they will be attributed to $\TEV$-scale physics and we can
obtain ``bottom-up'' type information about new physics. In this
paper, we consider the precision measurements of the top sector
at the LHC by using black holes. The stringent trigger conditions
to confirm the black hole production vanish almost all of the QCD
background, and we can examine the top quark emitted from black holes
very precisely. The error of the top quark mass and the top Yukawa coupling
are drastically reduced, leading to a very accurate test of the
Higgs mechanism. We can directly measure the CKM matrix element 
$|V_{ts}|$, and we will understand the property
of the CKM matrix and the origin of CP-violation deeply. 
The very precise measurements of such properties in the SM, enabled by
black holes, can become treasures in the quest for $\TEV$-scale
physics because there exists a possibility that 
$\TEV$-scale physics affects them and destroys the predictions of the SM. 
By combining the direct information of new physics 
obtained from black holes themselves and the 
indirect information obtained from the limitations of the SM, 
we will be able to identify $\TEV$-scale physics correctly.}

\end{center}
\end{titlepage}

\renewcommand{\thefootnote}{\arabic{footnote}}
\setcounter{footnote}{0}

%
%
%       *** Main Part ***
%
%

\section{Introduction}
\label{introduction}

The black hole production may be possible in $\TEV$-scale
gravity models at the $\TEV$-scale colliders, say at the LHC 
\cite{Giddings,Dimopoulos-Landsberg,LHC}, Tevatron \cite{Tevatron},
or linear colliders \cite{LC}.
High energy cosmic rays may also become the
origin of black holes \cite{Cosmicray}.
The properties of such black holes are 
extensively investigated \cite{Property}.
Since the black hole production processes
are purely gravitational and do not contain any small couplings,
their cross sections are very large and will be the first signal
of $\TEV$-scale gravity models.

So once black holes are produced, their primary roles are to give the
information for the quest of the true $\TEV$-scale physics.
The Standard Model (SM) will be recognized as the theory which describes
weak-scale physics, that is the low-energy approximation of
$\TEV$-scale physics.

But black holes have another potential. That is to test the
SM more precisely, and to examine the validity of the SM 
as the true weak-scale theory.

For example, if black holes are produced, Because of their
large masses, their decay products have high energies and
there is a possibility that new particles 
not found yet are produced by them.
\cite{Landsberg} investigated the prospect of discovering
Higgs bosons at the LHC by the decay products of black holes,
and concluded that only one day operation is enough to
discover Higgs bosons with $5 \sigma$ significance.
\cite{Uehara_spin} claimed that Higgs bosons produced 
by black holes enable us to measure
the spin of them at the LHC in only one month operation.
The spin measurement of Higgs bosons at the LHC was considered
to be almost impossible, so black holes offer us a new method
to measure the SM. 

The quest for the true $\TEV$-scale physics and the test
of the SM as weak-scale physics using black holes play
complementary role for the deeper understanding of many mechanisms
and phenomena which are
not completely understood yet, for example the origin of mass and
CP-violation. If only the SM is responsible for the origin of mass,
then the Higgs mechanism should work completely and any $\TEV$-scale
physics should not affect it. But there exists a possibility
that $\TEV$-scale physics is also responsible for the masses 
of some particles, and so we should test the SM as precisely as 
possible to obtain the information of the true $\TEV$-scale physics.

In this paper we consider the precision measurements of the top quarks 
produced by the decay of black holes at the LHC, 
and show that their properties are determined more precisely than 
usually considered. 

The error of the top quark mass is currently $5.1 \GEV$ \cite{CDFD0}. 
This is not enough value for the electroweak precision measurements. 
The masses of top, $W$ and Higgs are closely related in the SM, 
and thus the error reduction of the top quark mass will lead 
to the accurate confirmation of the SM as weak-scale physics.
All Yukawa coupling constants of fermions are not measured yet, 
and that of the top quark is expected
to be measured at the LHC. This direct measurement will make it clear
whether the Higgs mechanism is the origin of mass
or other mechanisms should be needed.
The measurements of CKM matrix elements 
are the necessity of the certification of the unitarity and 
the confirmation of the CKM matrix as the origin of CP-violation.
If the unitarity is broken, new physics like the fourth generation fermion 
is needed. And if the unitarity triangle turned out to be non-closed,
other mechanisms should be necessary to explain the observed
CP-violating processes.
LHC will measure $|V_{tb}|$ by the decay of 
many $t \bar{t}$-pairs produced by QCD processes.
Single top quarks produced by electroweak
processes, whose cross sections are proportional to $|V_{tb}|^{2}$,
can also be used for the measurement.

The utilization of top quarks produced by black holes can reduce
drastically the error of the top quark mass and the top Yukawa
constant. This reduction lead to a very precise test of the Higgs
mechanism. It also enables us to measure $|V_{ts}|$ directly,
which will be the precious information to confirm the unitarity
of the CKM matrix and to test whether the origin of CP-violation
is the CKM matrix only or not. They are all valuable high precision
tests of the SM as weak-scale physics.

The reason why black holes are useful for the precision
measurements of the top sector is that black holes can emit the
single, and high energy ($ \ge 1 \TEV$) top quark,
and the trigger conditions for the black hole production
make the QCD background almost be vanished. So we do not
have to struggle with the final state radiation, which
is the main source of the error of the top quark mass at the LHC.
The measurement of the top Yukawa coupling is released
from many background processes. The
strange quarks emitted from the processes $t \ra W s$
are not buried in huge QCD jets, and we can directly
measure $|V_{ts}|$ at the LHC.

Of course after the stringent trigger conditions which
almost vanish the huge and annoying QCD background,
the number of top quarks produced by black holes
are much less than the usually considered number of
$t \bar{t}$-pairs produced by QCD processes which 
amounts to 8 million pairs per year in low luminosity run of the LHC.
But the advantages from the trigger, namely the very
clean environment and the single and high energy top quark,
surpass the usual methods. So if $\TEV$-scale gravity
models are correct and black holes are produced at the
LHC, we should use the top quarks produced by black holes
for the deeper understanding of the top sector.
This understanding will lead to the stringent test of the SM,
and help us to make clear to what extent the SM is valid
as the theory which describes the weak-scale physics.

If we find an experimental result which cannot be explained
in the SM, it is attributed to $\TEV$-scale physics
and we can obtain much information about it. This ``bottom-up''
method should be used to identify new physics correctly.

This paper is organized as follows. In section \ref{BH},
we review the theories of $\TEV$-scale gravity and the
mechanism of the black hole production and decay.
After this preparation, we show the methods of using
top quarks produced by black holes for the precision measurements.
First, section \ref{mass} explains how to reduce the error
of the top quark mass. Next, section \ref{yukawa} is devoted
to the exposition of the precise measurement of the top Yukawa coupling.
Third, section \ref{CKM} investigates how to directly obtain the value
$|V_{ts}|$. We consider the unitarity of the CKM matrix and
whether CP-violation is attributed to the CKM matrix only or not.
Finally in section \ref{conclusion} we conclude.

\section{Black Hole Production and Decay}
\label{BH}

The possibility of $\TEV$-scale gravity was firstly proposed by
\cite{ADD,Antoniadis}. They proposed that the world is
$(4+n)$-dimensional and extra $n$ dimensions are compactified.
The observed Planck scale $M_{pl}$ is only valid for the four
dimensional spacetime, and the true Planck scale of the higher-dimensional
spacetime $M_{D}$ is given by:
\begin{eqnarray}
M_{D}^{n+2} V_{n} = M_{pl}^{2},
\end{eqnarray}
where $V_{n}$ is the compactified volume of the extra dimensions.
If we take $V_{n}$ properly, the true Planck scale can be 
${\rm O}(\TEV)$. So the hierarchy problem which arises from the
large difference between $M_{weak}$ and $M_{pl}$ 
no longer exists, and ${\rm O}(\TEV)$ is the true scale of
the quantum gravity.

If such $\TEV$-scale gravity models are correct, we can directly
access Planckian and transPlanckian region by real experiments.
The black hole production at the LHC was firstly pointed out by
\cite{Giddings,Dimopoulos-Landsberg}.
It is a very exciting phenomenon, and many papers investigated
their properties \cite{LHC}. Let us denote the mass of a black hole
by $M_{BH}$. Then the parton level cross section for the black hole
production is semiclassically given by \cite{Myers-Perry} :
\begin{eqnarray}
\sigma(M_{BH}) = \pi R_{S}^{2} = \frac{1}{M_{D}^{2}} \left[ \frac{M_{BH}}{M_{D}} (\frac{8 \Gamma(\frac{n+3}{2})}{n+2}) \right]^{2/(n+1)}, \label{BHeq}
\end{eqnarray}
where $R_{S}$ is the $(n+4)$-dimensional Schwarzshild radius.
After the convolution of parton distribution function of protons and the 
integration over $M_{BH}$, we obtain the total cross section.

But here we must be careful about the masses of black holes.
If $M_{BH} \sim M_{D}$ (Planckian black hole), 
the quantum gravity may drastically affect equation (\ref{BHeq}). 
Since nobody knows the true theory of the quantum gravity, 
we cannot use small value of $\frac{M_{BH}}{M_{D}}$. \cite{Cheung}
showed that when $\frac{M_{BH}}{M_{D}} \gsim 5$, the semiclassical
approximation is valid and equation (\ref{BHeq}) holds. So we take
the minimum value of the black hole mass to be $M_{BH}^{min} = 5 M_{D}$.

Next we have to determine $n$, the number of extra dimensions.
A review \cite{Uehara_review} showed that the cases of $n=2,3$ are
already excluded as the candidates of $\TEV$-scale gravity. and if we
take $n=4$, $M_{D} \gsim 2.3 \TEV$. Thus the energy of the LHC
$\sqrt{s} = 14 \TEV$ is the edge of the constraint 
$M_{BH} \gsim 5 M_{D}$, and the total cross section
becomes very small. In the cases of $n=5,6,7$, roughly $M_{D} \gsim 1 \TEV$
\cite{AGASALIMIT} and the total cross section becomes large enough. 
So we assume $n \gsim 5$ and $M_{D}=1 \TEV$.

Now we can calculate the total cross section. It becomes \cite{Rizzo}:
\begin{eqnarray}
\sigma = 10^{6} \ {\rm fb}. \label{sigma1}
\end{eqnarray}
Since the total cross section is suppressed exponentially
as we raise the masses of black holes, 
the main contribution for equation (\ref{sigma1}) comes from 
the black holes with masses $M_{BH} \sim 5 \TEV$.

But, there exists an argument that the production cross section for
transPlanckian black holes $(M_{BH} \gg M_{D})$
is suppressed by at least a factor 
$\exp(-I_{E})$ with $I_{E}$ being the Gibbons-Hawking action for
the black holes \cite{Voloshin}. Conservatively we adopt this
suppression factor. Then the total cross section is reduced to \cite{Rizzo}:
\begin{eqnarray}
\sigma = 10^{4} \ {\rm fb}.
\end{eqnarray}

Produced black holes decay by the Hawking radiation. The decay process
is governed by the temperature of black holes:
\begin{eqnarray}
T_{BH} = \frac{n+1}{4 \pi R_{S}},
\end{eqnarray}
and the spectrum of the black hole decay products is given by
averaging the Planck formula. By applying the Boltzmann statistics,
the mean energy carried by one emitted particle becomes:
\begin{eqnarray}
\vev{E} = 2 T_{BH}.
\end{eqnarray}
Thus the multiplicity $N$ of the produced black holes becomes:
\begin{eqnarray}
N = \frac{M_{BH}}{2 T_{BH}}.
\end{eqnarray}

As you can see from the figure 1(d) of \cite{Dimopoulos-Landsberg},
in the case of $5 \le n \le 7$ and $\frac{M_{BH}}{M_{D}}=5$,
$N$ is roughly 4. As stated above, most of the contribution
to the total cross section comes from the black holes
whose masses are $M_{BH} \sim 5 \TEV$. So approximately we obtain:
\bsub
\begin{eqnarray}
N &=& 4, \\
E &=& 1.3 \TEV,
\end{eqnarray}
\esub
where $E$ is the energy of each particle. Produced one black hole
emits four particles whose energies are about $1.3 \TEV$.

The decay of black holes do not discriminate any of the SM particles.
The probability of a certain particle being emitted from a black hole
depend on the degree of freedom of the particle. That of the SM
is about $120$.

We concentrate on the single top quark production. Thus
among the four particles emitted from one black hole, 
one particle must be the top quark and the others
should not contain top quarks. So as trigger conditions,
we require:
\begin{itemize}
 \item one jet and one lepton or two jets should exist
       inside the cone $\Delta R = 1.3$. (They are the decay products
       of the top quark.)
 \item Three particles or jets except top should be gluon, quarks except
       top, electron, muon or photon in order to certify four particles
       are surely emitted. 
 \item The total electric charge of observed particles Q must satisfy
       $|Q| \le \frac{4}{3}$. 
 \item Three particles or jets should have 
       $E_{T} \gsim 100 \GEV \sim 1300 \GEV \tm \sin 4^{\circ}$.
\end{itemize}
The degree of freedom of four particles which satisfies these
trigger conditions is roughly $1.1 \tm 10^{6}$. Thus the probability
that an event satisfies these trigger conditions becomes {\bf 0.0051}.

If we impose this trigger, the left QCD background with jets
have cross section less than $1 \ {\rm fb}$. 
(for a detail, see section 15 and 18 of \cite{ATLAS}). 
So the QCD background is completely negligible.

\section{Top Quark Mass}
\label{mass}

The usual top quark mass reconstruction process at the LHC utilizes the
$t \bar{t}$ pair production process, whose cross section is about
$\sigma(t \bar{t}) \sim 830 \ {\rm fb}$. They use the semileptonic
decay mode, $t \bar{t} \ra l \nu j j b \bar{b}$ for the reconstruction.
The result of \cite{ATLAS} is that in one year low luminosity run 
${\cal L} = 10 \ {\rm fb}^{-1}$, the error becomes:
\begin{eqnarray}
\delta m_{t} \sim 0.1 \GEV_{stat} \oplus 1.7 \GEV_{syst}.
\end{eqnarray}
So the systematic error dominates, and it cannot be easily reduced
even if the integrated luminosity is accumulated and the detector 
system is re-calibrated.

The main source of the systematic error is the presence of final
state radiation (FSR). The FSR affects the reconstructed top quark mass
directly since the reconstructed top quark mass receives
the effect of the jets which are radiated and escaped from the cone.
It is clearly shown in table 18-3 of \cite{ATLAS}.

One way to escape from this effect is to increase the cone size.
If we increase the cone size to $\Delta R = 1.3$, the effects
of the FSR are almost vanished. But as a compensation, the underlying
events will enter in the expanded cone and the reconstruction 
process of $W$ from jet-pair suffer from background QCD multi-jet
events. As a result, we can surely reduce the systematic error
by increasing the cone size, $\delta m_{syst} \sim 1.7 \GEV$ is
the lower bound at this time. (see table 18-4 of \cite{ATLAS}.)

From now we explain the advantage of our method. From section \ref{BH}, 
the number of single top quark which are produced by black holes
becomes:
\begin{eqnarray}
10^{4} \ {\rm fb} \tm 10^{2} \ {\rm fb}^{-1} \tm 0.0051 = 5100.
\end{eqnarray}
We use the $t \ra j j b$ and $t \ra l \nu b$ modes both.
From the discussion of section \ref{BH}, the QCD background is
almost vanished. So we do not have to worry about the underlying
events. Even if we set the large cone size, namely $\Delta R = 1.3$,
no problem occurs. Thus the systematic error is now dominated by
the uncertainty of cell energy scale. In our case it is roughly
$\delta m_{t} = 0.5 \GEV$. Now the systematic error is considerably
reduced.

Next consider about the statistical error. we can calculate it 
straightforwardly. 
From \cite{ATLAS}, the hadronic mode's statistical error
becomes:
\begin{eqnarray}
0.070 \tm \sqrt{\frac{32000}{5100 \tm 0.6 \tm (6/9)}} = 0.28 \GEV,
\end{eqnarray}
where $0.6$ is the b-tagging efficiency in low luminosity run,
and $(6/9)$ is the branching ratio of the hadronic mode. 
And the leptonic mode's statistical error becomes:
\begin{eqnarray}
0.9 \tm \sqrt{\frac{15200 \tm 0.85}{5100 \tm 0.6 \tm (2/9)}} = 3.9 \GEV,
\end{eqnarray}
where $0.85$ is the ratio which describes the correctness of
$lb$ pairing in the $t \bar{t}$ dilepton decay mode, and $(2/9)$
is the branching ratio of the leptonic mode.

So the leptonic mode suffers from the large statistical error, and
we do not adopt this mode in the final result. Our final result is that,
with the integrated luminosity $10 \ {\rm fb}^{-1}$ which can be
accumulated in one year low luminosity run of the LHC, the error
of the top quark mass becomes:
\begin{eqnarray}
\Delta m_{t} = 0.28 \GEV_{stat} \oplus 0.5 \GEV_{syst}.
\end{eqnarray}
Although the statistical error increases because of the small number
of top quarks, the extreme clean environment made from the black hole production
and its trigger conditions reduced the systematic error drastically,
and leads to the small total error $\Delta m_{t} = 0.57 \GEV.$

Now we summarize the (expected) error of the top quark mass.
\bsub
\begin{eqnarray}
\Delta m_{t} &=& 3.2 \GEV_{stat} \oplus 4.0 \GEV_{syst} = 5.1 \GEV \ ({\rm CDF+D0 \ combined }), \label{mteq1} \\ 
\Delta m_{t} &=& 3 \GEV \ ({\rm Run \ IIa \ of \ the \ Tevatron}), \label{mteq2} \\
\Delta m_{t} &=& 2 \GEV \ ({\rm Run \ IIb \ of \ the \ Tevatron}), \label{mteq3} \\
\Delta m_{t} &=& 0.1 \GEV_{stat} \oplus 1.7 \GEV_{syst} = 1.7 \GEV \ ({\rm the \ LHC, \ 10 \ fb^{-1}}), \\
\Delta m_{t} &=& 0.28 \GEV_{stat} \oplus 0.5 \GEV_{syst} = 0.57 \GEV \ ({\rm the \ LHC, \ 10 \ fb^{-1}, \ using \ black \ holes}), \nonumber \\
\end{eqnarray}
\esub
where (\ref{mteq1}) is the latest result from Tevatron \cite{CDFD0}
(current error), (\ref{mteq2}) and (\ref{mteq3}) are the expected errors 
in the Tevatron Run IIa \cite{Run2a} and Run IIb \cite{Run2b}.

As stated in the section \ref{introduction}, the masses of top, $W$
and Higgs are closely related in the SM. It is shown in figure \ref{mtmwfig}.
The LHC can measure the $W$ boson mass with the uncertainty 
$\Delta m_{W} = 25 \MEV$ \cite{ATLAS}. From the figure \ref{mtmwfig},
we observe that $\Delta m_{t}=0.59 \GEV$ is a very satisfactory value
in order to determine the Higgs boson mass from this relationship.
We should reduce the error of the $W$ boson mass in order to make
the prediction for the Higgs boson mass more accurately, and to
test the SM more precisely.

\begin{figure}[htbp]
\centerline{\psfig{file=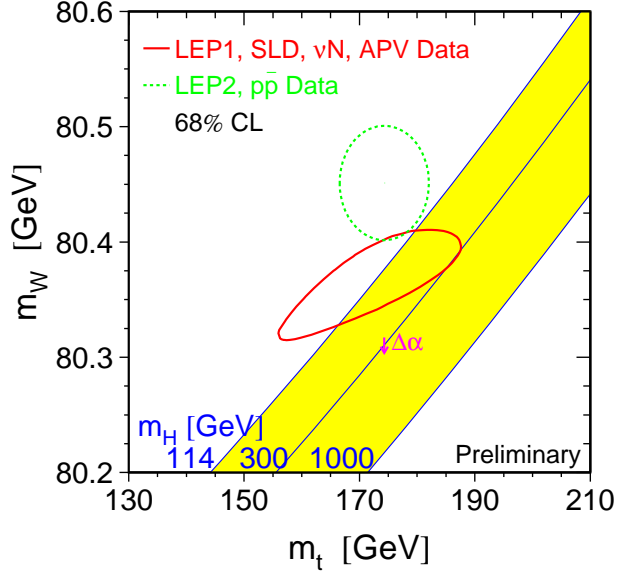,height=9cm}}
\caption{The latest result from LEP and SLD \protect\cite{LEP}.}
\label{mtmwfig}
\end{figure}

\section{Top Yukawa Coupling}
\label{yukawa}

Every fermions except neutrinos obtain their masses by the
Higgs mechanism, and their masses are proportional to their
Yukawa couplings. But up to now no Yukawa couplings are
directly measured. The direct measurement of the top Yukawa coupling
will clearfy whether the Higgs mechanism is the true origin of
mass or not.

At the LHC, the top Yukawa coupling can be determined by the
$t \bar{t} H$ production mode \cite{ATLAS}. The analysis of this event
require one of the top quark decay leptonically, and another one
decay hadronically. They assume the Higgs boson mass $m_{h}=120 \GEV$,
and use the main decay mode $h \ra b \bar{b}$. So the final state
contains four b-jets, leading to the large combinatorial background.
Many of the systematic errors, such as those associated with
uncertainties in the integrated luminosity and in the $t \bar{t}$
reconstruction efficiency, could be controlled by comparing the 
$t \bar{t} H$ rate with the $t \bar{t}$ rate.

Now we consider the usage of the single top quark produced by
black holes. Most of the top quarks decay into $Wb$, but
there exists a mode that the top quark emits a Higgs boson,
namely $t \ra t H$. The branching ratio of this mode becomes:
\begin{eqnarray}
\mbox{Br} (t \ra t H) = 0.046 \frac{y_{t}^{2}}{|V_{tb}|^{2}} \sim 0.046 y_{t}^{2}.
\end{eqnarray}
Here we used the fact $|V_{tb}| \sim 1$.

Once a Higgs boson is emitted, it immediately decays. The Higgs 
boson whose mass is $m_{h}=120 \GEV$ has the following branching
ratios \cite{HDECAY}:
\bsub
\begin{eqnarray}
{\rm Br} (h  \ra  b \bar{b}) &\sim& 0.66, \label{heq1} \\
{\rm Br} (h  \ra  W W) &\sim& 0.14, \\
{\rm Br} (h  \ra  g  g) &\sim& 0.076, \\
{\rm Br} (h  \ra  \tau \bar{\tau}) &\sim& 0.074. \label{heq4}
\end{eqnarray}
\esub
For the reconstruction of Higgs bosons, we use the above modes.
The decay products of Higgs bosons have energies
well less than those of other three particles (or jets), and
from the trigger conditions, the existence of one top quark
is assured. And the QCD background is almost vanished after
the trigger conditions. 

So we do not have to require for b-quarks to be b-tagged. 
Gluons immediately hadronize and the reconstruction is straightforward.
Hadronic, semileptonic and leptonic decay modes of $WW$ and 
$\tau \bar{\tau}$ pair can be used for the reconstruction.
Including all of the available processes, the reconstruction efficiency
for the $120 \GEV$ Higgs bosons becomes about $0.9$. 

So with the integrated luminosity ${\cal L}=30 \ {\rm fb}^{-1}$,
The number of reconstructed Higgs bosons becomes:
\begin{eqnarray}
5100 &\tm& 0.6 \ (\mbox{b-tag efficiency for the top quark reconstruction}) \nonumber \\ 
&\tm& (2/9 + 6/9) \ (\mbox{W reconstruction efficiency}) \nonumber \\
&\tm& 3 \ (\mbox{30 fb factor}) \tm 0.046 y_{t}^{2} \tm 0.9 \nonumber \\
&=& 340 y_{t}^{2}.
\end{eqnarray}
If we assume the integrated luminosity ${\cal L} = 100 \ {\rm fb}^{-1}$,
the b-tag efficiency is reduced to $0.5$, and the number becomes:
\begin{eqnarray}
940 y_{t}^{2}.
\end{eqnarray}
From the obtained data, we can estimate the expected error of
the top Yukawa coupling. It is shown in table \ref{topyukawatab}.
\btable
\begin{tabular}{|r|c|c|}
\hline
integrated luminosity & 30 ${\rm fb}^{-1}$ & 100 ${\rm fb}^{-1}$ \\
\hline
without black holes & $16.2 \%$ & $14.4 \%$ \\
\hline
with black holes & $2.7 \%$ & $1.6 \%$ \\
\hline
\etable{The expected statistical error of the top Yukawa
 coupling at the LHC. \label{topyukawatab}}
The extreme clean environment enable us to reduce the
error drastically. From section \ref{mass} we observe
that the error of top quarks at $30 \ {\rm fb}^{-1}$ becomes
$0.5 \GEV$, which corresponds to $0.3 \%$ error.
The measurements of the top quark mass and the top Yukawa coupling
with such a high accuracy will surely make it clear whether
the Higgs mechanism, the most important and crucial 
part in the SM, is true or not. If some discrepancies from the
prediction of the Higgs mechanism are found, they imply that
the SM only cannot explain the origin of mass, and 
some mechanisms which arise from $\TEV$-scale physics 
are necessary for the correct understanding of nature of mass.

\section{CKM matrix element $|V_{ts}|$}
\label{CKM}

The measurements of the CKM matrix elements will clearfy the 
origin of the observed CP violation,
and they will also make it clear whether the unitarity of the CKM matrix
holds or not. The high statistics of LHC $t \bar{t}$-pairs
(8 million pairs per year in low luminosity run) can measure 
$|V_{tb}|$ by the decay $t \ra W b$ with the accuracy $0.2 \%$
(statistical error only) \cite{ATLAS}.

Since in our setup the number of top quarks is much less than
8 million, we cannot compete with this precision. But
in our case the QCD background is almost vanished, we can measure
the other CKM matrix elements like $|V_{ts}|$. Here we concentrate
on the measurement of $|V_{ts}|$.

From the fact $|V_{td}| \ll |V_{ts}| \ll |V_{tb}|$, the branching
ratio of $t \ra W s$ is simply given by:
\begin{eqnarray}
\mbox{Br} (t \ra W s) = |\frac{V_{ts}}{V_{tb}}|^{2} \sim |V_{ts}|^{2}.
\end{eqnarray}
Here we used $|V_{tb}| \sim 1$.

The strange quarks emitted from top quarks form mesons like
$K^{\pm}$ or $K^{0}$. Their mean decay lengths are measured to be \cite{PDG}:
\bsub
\begin{eqnarray}
K^{\pm} &:& 3.713 \ \mbox{m}, \\
K_{S}^{0} &:& 2.6786 \ \mbox{cm}, \\
K_{L}^{0} &:& 15.51 \ \mbox{m}.
\end{eqnarray}
\esub
Thus $K_{S}^{0}$ decay inside the inner detector. But $K^{\pm}$ and
$K_{L}^{0}$ penetrate it and detected at the electromagnetic calorimeter
(if the formed K meson is charged) and the hadronic calorimeter.

Bottom mesons decay inside the inner detector. So in this
clean environment, we can distinguish $t \ra W b$ and $t \ra W s$
if $W$ decays leptonically and $K_{S}^{0}$ is not formed, by
examining the number of layers which detected some events 
in the inner detector.
Thus the reconstruction efficiency of $t \ra W s$ mode becomes:
\begin{eqnarray}
\sim 2/9 \tm 0.75 = 0.17.
\end{eqnarray}
And so the number of events $t \ra W s$ with the integrated
luminosity ${\cal L}=10 \ \mbox{fb}^{-1}$ is given by:
\begin{eqnarray}
5100 \tm |V_{ts}|^{2} \tm 0.17 = 870 |V_{ts}|^{2}. \label{Vtseq}
\end{eqnarray}
The current value of $|V_{ts}|$ is, at the $90 \%$ confidence level \cite{PDG}:
\begin{eqnarray}
0.037 \le |V_{ts}| \le 0.043.
\end{eqnarray}
If we substitute $|V_{ts}|=0.04$ into (\ref{Vtseq}), only $1.4$ events
are expected. This is not at all enough value to determine 
$|V_{ts}|$ or its error. So let us assume the integrated luminosity 
${\cal L} = 300 \ \mbox{fb}^{-1}$.
Then $42$ events are expected and we can directly determine $|V_{ts}|$ 
(note that the current value is the bound obtained from the other
information) and the error of $|V_{ts}|$.

The integrated luminosity ${\cal L} = 300 \mbox{fb}^{-1}$ enables us
to measure $|V_{ts}|$ with the accuracy:
\begin{eqnarray}
8.3 \% \ (\mbox{in the case} \ |V_{ts}|=0.037) \sim 7.1 \% \ (\mbox{in the case} \ |V_{ts}|=0.043).
\end{eqnarray}
This result will greatly help us to determine whether the unitarity
of the CKM matrix is correct or not, CP-violating processes are correctly
described by the CKM matrix or not, and so on.

The unitarity can easily be checked by examining whether the following
equation holds or not:
\begin{eqnarray}
|V_{td}|^{2} + |V_{ts}|^{2} + |V_{tb}|^{2} = 1. \label{unitarityeq}
\end{eqnarray}
$|V_{tb}|$ can be measured at the LHC with very large $t \bar{t}$
samples and its error is expected to be very small. And the value of
$|V_{td}|$ is negligibly small. So the direct measurement of $|V_{ts}|$
will enable us to verify whether equation (\ref{unitarityeq}), namely
the unitarity condition, is correct or not.

Next consider about CP-violation. If the unitarity triangle 
(see figure \ref{trianglefig}) is
not closed, CP-violation cannot be explained by the CKM matrix only.
When we use the Wolfenstein
approximation of the triangle \cite{Wolfenstein}, the CKM matrix is
parameterized by $A, \ \rho, \ \mbox{and} \ \eta$. And we have the
relationship:
\begin{eqnarray}
\frac{|V_{td}|^{2}}{|V_{ts}|^{2}} = (1-\rho)^{2} + \eta^{2}. \label{rhoetaeq}
\end{eqnarray}
Usually we can use this relationship by the observables 
$\Delta M_{B_{d}}$ and $\Delta M_{B_{s}}$, since $|V_{td}|$ and
$|V_{ts}|$ are not directly measured yet. Furthermore, currently
only a lower bound on $\Delta M_{B_{s}}$ is measured. So the current 
constraint which 
depends on this relationship can only set an upper bound.

\begin{figure}[htbp]
\centerline{\psfig{file=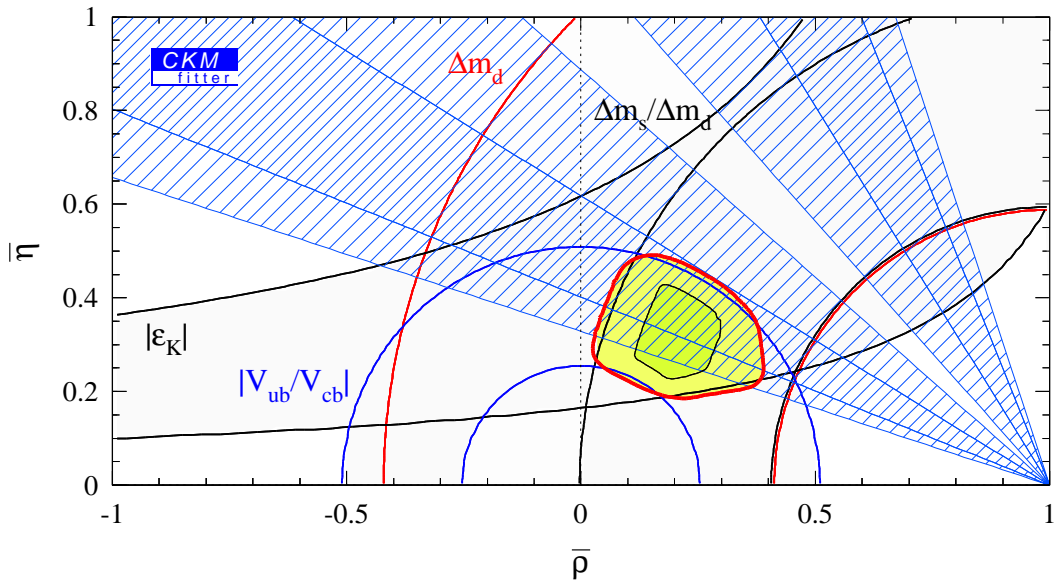,height=8cm}}
\caption{The latest summary of unitarity triangle constraints with the
 direct measurements of $\sin (2 \beta)$ by Belle \protect\cite{Belle}
 : $\sin (2 \beta) = 0.99 \pm 0.15$ and Babar \protect\cite{Babar} : 
 $\sin (2 \beta) = 0.59 \pm 0.15$. (From \cite{Beneke,Hocker})}
\label{trianglefig}
\end{figure}

But assume that
we have directly measured $|V_{ts}|$ at the LHC by using black holes.
Then equation (\ref{rhoetaeq}) becomes:
\begin{eqnarray}
(1-\rho)^{2} + \eta^{2} = \frac{1 - |V_{tb}|^{2} - |V_{ts}|^{2}}{|V_{ts}|^{2}}.
\end{eqnarray}
Here we assumed the unitarity condition.
Now we can constrain the unitarity triangle by the experimentally obtained
data only, which cannot be possible when we use the ratio of
$\Delta M_{B_{d}}$ to $\Delta M_{B_{s}}$. This is because they contain
some parameters which describes the hadronization possibilities and
such parameters cannot be experimentally accessible.
And furthermore, we can also set a lower bound on the radius of the circle.

Finally consider about the rare decay mode $b \ra s \gamma$. Its inclusive
decay width is given by \cite{Stone}:
\begin{eqnarray}
\Gamma (b \ra s \gamma) = \frac{G_{F}^{2} \alpha m_{b}^{5}}{32 \pi^{4}} |C_{7}|^{2} |V_{tb} V_{ts}^{*}|^{2}.
\end{eqnarray}
The current theoretical prediction of the branching ratio 
$\mbox{Br} (b \ra s \gamma)$ in the SM is 
$(3.73 \pm 0.30) \tm 10^{-4}$ \cite{Hurth}, and the current
experimental average is $(3.23 \pm 0.42) \tm 10^{-4}$ \cite{Stone}.
So the error reduction
of $|V_{ts}|$ will also reduce the theoretical error of 
$\mbox{Br}(b \ra s \gamma)$, and
we may observe whether the $b \ra s \gamma$ process can be
explained by the SM only, or some mechanisms arise from 
new physics are needed.

\section{Conclusion}
\label{conclusion}

In this paper, we consider the possibility of measuring the top
sector in the SM more precisely at the LHC, by using
black holes which will be produced if $\TEV$-scale gravity models
are correct. From the stringent trigger conditions to confirm
black holes are certainly produced, the huge QCD background which
obstruct the precision measurements is almost vanished. 
From the single top quark which is produced by the decay of black holes,
we can measure its mass, Yukawa coupling and the CKM matrix element
$|V_{ts}|$ precisely.

The error of the top quark mass is reduced from $1.7 \GEV$ to
$0.56 \GEV$, because the systematic error originated from the
final state radiation can be evaded by the extreme clean environment.
The error of the top Yukawa coupling is also reduced about a factor 
$6 \sim 9$. The mass and the Yukawa coupling of the top quark are the
crucial information for the clarification of the origin of mass.
The SM relies on the Higgs mechanism, but this is not examined yet
by the real experiment. By using black holes, we can test the Higgs
mechanism with a very high accuracy.

The CKM matrix element $|V_{tb}|$ can be measured at the LHC
by using many $t \bar{t}$-pairs produced by QCD processes.
In our case, the extreme clean environment makes it possible to measure
$|V_{ts}|$ directly. This data is very important
to test whether the origin of CP-violation is attributed to
the CKM matrix only or not,
and to validate the unitarity of the CKM matrix.

When we find an experimental result which is inconsistent with the mechanism
in the SM, it implies that some new mechanisms arise from $\TEV$-scale
physics should be responsible for the result. The masses of fermions,
some CP-violating processes or the fourth generation for example, 
may be attributed to $\TEV$-scale physics. So 
the precision measurements of the SM is deeply connected to the
quest for new physics which governs $\TEV$-scale.

To summarize, once black holes are produced at the LHC, we will recognize
the end of the SM as the theory which can describe $\TEV$-scale
and the search for the true $\TEV$-scale physics becomes
the problem of the utmost importance. But on the other hand,
black holes can also be used to test to what extent the SM is correct.
The search for the new physics and the precision measurements of the SM 
using black holes will play complementary roles for the deep understandings
of many mechanisms and phenomena not uncovered yet.

\vt

{\bf Acknowledgment}

Y.U. thank Japan Society for the Promotion of Science for financial
support.

\vt


\begin{thebibliography}{99}
%

\bibitem{Giddings}
S.~B.~Giddings and S.~Thomas,
%``High energy colliders as black hole factories: The end of short  distance physics,''
Phys.\ Rev.\ D {\bf 65}, 056010 (2002)
[arXiv:hep-ph/0106219].
%%CITATION = HEP-PH 0106219;%%

\bibitem{Dimopoulos-Landsberg}
S.~Dimopoulos and G.~Landsberg,
%``Black holes at the LHC,''
Phys.\ Rev.\ Lett.\  {\bf 87}, 161602 (2001)
[arXiv:hep-ph/0106295].
%%CITATION = HEP-PH 0106295;%%


\bibitem{LHC}
S.~Dimopoulos and R.~Emparan,
%``String balls at the LHC and beyond,''
Phys.\ Lett.\ B {\bf 526}, 393 (2002)
[arXiv:hep-ph/0108060].
%%CITATION = HEP-PH 0108060;%%
S.~Hossenfelder, S.~Hofmann, M.~Bleicher and H.~Stocker,
%``Quasi-stable black holes at LHC,''
arXiv:hep-ph/0109085.
%%CITATION = HEP-PH 0109085;%%
S.~Hofmann, M.~Bleicher, L.~Gerland, S.~Hossenfelder, S.~Schwabe and H.~Stocker,
%``Suppression of high-P(T) jets as a signal for large extra dimensions
% and new estimates of lifetimes for meta stable micro black holes:  From 
% the early universe to future colliders,''
arXiv:hep-ph/0111052.
%%CITATION = HEP-PH 0111052;%%
S.~C.~Park and H.~S.~Song,
%``Production of spinning black holes at colliders,''
arXiv:hep-ph/0111069.
%%CITATION = HEP-PH 0111069;%%
G.~F.~Giudice, R.~Rattazzi and J.~D.~Wells,
%``Transplanckian collisions at the LHC and beyond,''
Nucl.\ Phys.\ B {\bf 630}, 293 (2002)
[arXiv:hep-ph/0112161].
%%CITATION = HEP-PH 0112161;%%


\bibitem{Tevatron}
M.~Bleicher, S.~Hofmann, S.~Hossenfelder and H.~Stocker,
%``Black hole production in large extra dimensions at the Tevatron: New  limit on the fundamental scale of gravity,''
arXiv:hep-ph/0112186.
%%CITATION = HEP-PH 0112186;%%

\bibitem{LC}
Y.~Uehara,
%``Virtual Black Holes at Linear Colliders,''
arXiv:hep-ph/0205068.
%%CITATION = HEP-PH 0205068;%%



\bibitem{Cosmicray}
J.~L.~Feng and A.~D.~Shapere,
%``Black hole production by cosmic rays,''
Phys.\ Rev.\ Lett.\  {\bf 88}, 021303 (2002)
[arXiv:hep-ph/0109106].
%%CITATION = HEP-PH 0109106;%%
%``Experimental signature for black hole production in neutrino air  showers,''
Phys.\ Rev.\ D {\bf 65}, 047502 (2002)
[arXiv:hep-ph/0109242].
%%CITATION = HEP-PH 0109242;%%
R.~Emparan, M.~Masip and R.~Rattazzi,
%``Cosmic rays as probes of large extra dimensions and TeV gravity,''
Phys.\ Rev.\ D {\bf 65}, 064023 (2002)
[arXiv:hep-ph/0109287].
%%CITATION = HEP-PH 0109287;%%
.~Kazanas and A.~Nicolaidis,
%``Cosmic rays and large extra dimensions,''
arXiv:hep-ph/0109247.A.~Ringwald and H.~Tu,
%``Collider versus cosmic ray sensitivity to black hole production,''
Phys.\ Lett.\ B {\bf 525}, 135 (2002)
[arXiv:hep-ph/0111042].
%%CITATION = HEP-PH 0111042;%%
Y.~Uehara,
%``Production and detection of black holes at neutrino array,''
Prog.\ Theor.\ Phys.\  {\bf 107}, 621 (2002)
[arXiv:hep-ph/0110382].
%%CITATION = HEP-PH 0110382;%%
A.~Ringwald and H.~Tu,
%``Collider versus cosmic ray sensitivity to black hole production,''
Phys.\ Lett.\ B {\bf 525}, 135 (2002)
[arXiv:hep-ph/0111042].
%%CITATION = HEP-PH 0111042;%%
M.~Kowalski, A.~Ringwald and H.~Tu,
%``Black holes at neutrino telescopes,''
Phys.\ Lett.\ B {\bf 529}, 1 (2002)
[arXiv:hep-ph/0201139].
%%CITATION = HEP-PH 0201139;%%
P.~Jain, S.~Kar, S.~Panda and J.~P.~Ralston,
%``Brane-production and the neutrino nucleon cross section at ultra high  energies in low scale gravity models,''
arXiv:hep-ph/0201232.
%%CITATION = HEP-PH 0201232;%%
J.~Alvarez-Muniz, J.~L.~Feng, F.~Halzen, T.~Han and D.~Hooper,
%``Detecting microscopic black holes with neutrino telescopes,''
arXiv:hep-ph/0202081.
%%CITATION = HEP-PH 0202081;%%
L.~A.~Anchordoqui, J.~L.~Feng and H.~Goldberg,
%``p-branes and the GZK paradox,''
arXiv:hep-ph/0202124.
%%CITATION = HEP-PH 0202124;%%
J.~J.~Friess, T.~Han and D.~Hooper,
%``TeV string state excitation via high energy cosmic neutrinos,''
arXiv:hep-ph/0204112.
%%CITATION = HEP-PH 0204112;%%
S.~I.~Dutta, M.~H.~Reno and I.~Sarcevic,
%``On black hole detection with the OWL/Airwatch telescope,''
arXiv:hep-ph/0204218.
%%CITATION = HEP-PH 0204218;%%
H.~Tu,
%``Microscopic Black Hole Production in TeV-Scale Gravity,''
arXiv:hep-ph/0205024.
%%CITATION = HEP-PH 0205024;%%


\bibitem{Property}
P.~C.~Argyres, S.~Dimopoulos and J.~March-Russell,
%``Black holes and sub-millimeter dimensions,''
Phys.\ Lett.\ B {\bf 441}, 96 (1998)
[arXiv:hep-th/9808138].
%%CITATION = HEP-TH 9808138;%%
H.~C.~Kim, S.~H.~Moon and J.~H.~Yee,
%``Dimensional dependence of black hole formation in scalar field  collapse,''
JHEP {\bf 0202}, 046 (2002)
[arXiv:gr-qc/0108071].
%%CITATION = GR-QC 0108071;%%
R.~Casadio and B.~Harms,
%``Can black holes and naked singularities be detected in accelerators?,''
arXiv:hep-th/0110255.
%%CITATION = HEP-TH 0110255;%%
D.~M.~Eardley and S.~B.~Giddings,
%``Classical black hole production in high-energy collisions,''
arXiv:gr-qc/0201034.
%%CITATION = GR-QC 0201034;%%
E.~J.~Ahn, M.~Cavaglia and A.~V.~Olinto,
%``Brane factories,''
arXiv:hep-th/0201042.
%%CITATION = HEP-TH 0201042;%%
S.~N.~Solodukhin,
%``Classical and quantum cross-section for black hole production in  particle collisions,''
Phys.\ Lett.\ B {\bf 533}, 153 (2002)
[arXiv:hep-ph/0201248].
%%CITATION = HEP-PH 0201248;%%
V.~Cardoso and J.~P.~Lemos,
%``Gravitational radiation from collisions at the speed of light: A  massless particle falling into a Schwarzschild black hole,''
arXiv:gr-qc/0202019.
%%CITATION = GR-QC 0202019;%%
E.~Kohlprath and G.~Veneziano,
%``Black holes from high-energy beam-beam collisions,''
arXiv:gr-qc/0203093.
%%CITATION = GR-QC 0203093;%%
P.~Kanti and J.~March-Russell,
%``Calculable corrections to brane black hole decay. I: The scalar case,''
arXiv:hep-ph/0203223.
%%CITATION = HEP-PH 0203223;%%
S.~D.~Hsu,
%``Quantum production of black holes,''
arXiv:hep-ph/0203154.
%%CITATION = HEP-PH 0203154;%%
A.~V.~Kotwal and S.~Hofmann,
%``Discrete energy spectrum of Hawking radiation from Schwarzschild  surfaces,''
arXiv:hep-ph/0204117.
%%CITATION = HEP-PH 0204117;%%
L.~A.~Anchordoqui, H.~Goldberg and A.~D.~Shapere,
%``Phenomenology of Randall-Sundrum black holes,''
arXiv:hep-ph/0204228.
%%CITATION = HEP-PH 0204228;%%
K.~Cheung,
%``Black hole, string ball, and p-brane production at hadronic  supercolliders,''
arXiv:hep-ph/0205033.
%%CITATION = HEP-PH 0205033;%%
E.~J.~Ahn and M.~Cavaglia,
%``A New Era in High-energy Physics,''
arXiv:hep-ph/0205168.
%%CITATION = HEP-PH 0205168;%%





\bibitem{Landsberg}
G.~Landsberg,
%``Discovering new physics in the decays of black holes,''
Phys.\ Rev.\ Lett.\  {\bf 88}, 181801 (2002)
[arXiv:hep-ph/0112061].
%%CITATION = HEP-PH 0112061;%%


\bibitem{Uehara_spin}
Y.~Uehara,
%``Black Holes at the LHC can Determine the Spin of Higgs Bosons,''
arXiv:hep-ph/0205122.
%%CITATION = HEP-PH 0205122;%%

\bibitem{CDFD0}
G.~Brooijmans  [CDF and D0 Collaborations],
%``Top quark mass measurements at the Tevatron,''
arXiv:hep-ex/0005030.
%%CITATION = HEP-EX 0005030;%%

\bibitem{ADD}
N.~Arkani-Hamed, S.~Dimopoulos and G.~R.~Dvali,
%``The hierarchy problem and new dimensions at a millimeter,''
Phys.\ Lett.\ B {\bf 429}, 263 (1998)
[arXiv:hep-ph/9803315].
%%CITATION = HEP-PH 9803315;%%

\bibitem{Antoniadis}
I.~Antoniadis, N.~Arkani-Hamed, S.~Dimopoulos and G.~R.~Dvali,
%``New dimensions at a millimeter to a Fermi and superstrings at a TeV,''
Phys.\ Lett.\ B {\bf 436}, 257 (1998)
[arXiv:hep-ph/9804398].
%%CITATION = HEP-PH 9804398;%%


\bibitem{Myers-Perry}
R.~C.~Myers and M.~J.~Perry,
%``Black Holes In Higher Dimensional Space-Times,''
Annals Phys.\  {\bf 172}, 304 (1986).
%%CITATION = APNYA,172,304;%%




\bibitem{Cheung}
K.~m.~Cheung,
%``Black hole production and large extra dimensions,''
arXiv:hep-ph/0110163.
%%CITATION = HEP-PH 0110163;%%

\bibitem{Uehara_review}
Y.~Uehara,
%``A mini-review of constraints on extra dimensions,''
arXiv:hep-ph/0203244.
%%CITATION = HEP-PH 0203244;%%

\bibitem{AGASALIMIT}
L.~A.~Anchordoqui, J.~L.~Feng, H.~Goldberg and A.~D.~Shapere,
%``Black holes from cosmic rays: Probes of extra dimensions and new limits  on TeV-scale gravity,''
arXiv:hep-ph/0112247.
%%CITATION = HEP-PH 0112247;%%



\bibitem{Rizzo}
T.~G.~Rizzo,
%``Black hole production at the LHC: Effects of Voloshin suppression,''
JHEP {\bf 0202}, 011 (2002)
[arXiv:hep-ph/0201228].
%%CITATION = HEP-PH 0201228;%%


\bibitem{Voloshin}
M.~B.~Voloshin,
%``Semiclassical suppression of black hole production in particle  collisions,''
Phys.\ Lett.\ B {\bf 518}, 137 (2001)
[arXiv:hep-ph/0107119].
%%CITATION = HEP-PH 0107119;%%
M.~B.~Voloshin,
%``More remarks on suppression of large black hole production in particle  collisions,''
Phys.\ Lett.\ B {\bf 524}, 376 (2002)
[arXiv:hep-ph/0111099].
%%CITATION = HEP-PH 0111099;%%


\bibitem{ATLAS}
ATLAS Detector and Physics Performance
Technical Design Report, CERN-LHCC 99-14. 

\bibitem{Run2a}
R.~Blair {\it et al.}  [CDF-II Collaboration], ``The CDF-II detector: Technical
design report,'' FERMILAB-PUB-96-390-E.
S.~Abachi {\it et al.} [The D0 Collaboration], ``The D0 upgrade: The detector
and its physics,'' FERMILAB-PUB-96-357-E.

\bibitem{Run2b}
[TeV-2000 Study Group Collaboration], ``Future electroweak physics at the
Fermilab Tevatron: Report of the TeV-2000 Study Group,'' FERMILAB-PUB-96-082.

\bibitem{LEP}
D.~Abbaneo {\it et al.}  [ALEPH Collaboration],
%``A combination of preliminary electroweak measurements and constraints  on the standard model,''
arXiv:hep-ex/0112021.
%%CITATION = HEP-EX 0112021;%%


\bibitem{HDECAY}
A.~Djouadi, J.~Kalinowski and M.~Spira,
%``HDECAY: A program for Higgs boson decays in the standard model and its  supersymmetric extension,''
Comput.\ Phys.\ Commun.\  {\bf 108}, 56 (1998)
[arXiv:hep-ph/9704448].
%%CITATION = HEP-PH 9704448;%%

\bibitem{PDG}
D.~E.~Groom {\it et al.}  [Particle Data Group Collaboration],
%``Review Of Particle Physics,''
Eur.\ Phys.\ J.\ C {\bf 15}, 1 (2000).
%%CITATION = EPHJA,C15,1;%%


\bibitem{Wolfenstein}
L.~Wolfenstein,
%``Parametrization Of The Kobayashi-Maskawa Matrix,''
Phys.\ Rev.\ Lett.\  {\bf 51}, 1945 (1983).
%%CITATION = PRLTA,51,1945;%%

\bibitem{Belle}
K.~Abe {\it et al.}  [Belle Collaboration],
%``Observation of large CP violation in the neutral B meson system,''
Phys.\ Rev.\ Lett.\  {\bf 87}, 091802 (2001)
[arXiv:hep-ex/0107061].
%%CITATION = HEP-EX 0107061;%%


\bibitem{Babar}
B.~Aubert {\it et al.}  [BABAR Collaboration],
%``Observation of CP violation in the B0 meson system,''
Phys.\ Rev.\ Lett.\  {\bf 87}, 091801 (2001)
[arXiv:hep-ex/0107013].
%%CITATION = HEP-EX 0107013;%%


\bibitem{Beneke}
M.~Beneke,
%``CP violation,''
arXiv:hep-ph/0201137.
%%CITATION = HEP-PH 0201137;%%

\bibitem{Hocker}
A.~Hocker, H.~Lacker, S.~Laplace and F.~Le Diberder,
%``A new approach to a global fit of the CKM matrix,''
Eur.\ Phys.\ J.\ C {\bf 21}, 225 (2001)
[arXiv:hep-ph/0104062].
%%CITATION = HEP-PH 0104062;%%


\bibitem{Stone}
S.~Stone,
%``B phenomenology,''
arXiv:hep-ph/0112008.
%%CITATION = HEP-PH 0112008;%%


\bibitem{Hurth}
T.~Hurth,
%``Inclusive rare B decays,''
in {\it Proc. of the 5th International Symposium on Radiative Corrections (RADCOR 2000) } ed. Howard E. Haber,
arXiv:hep-ph/0106050.
%%CITATION = HEP-PH 0106050;%%


\end{thebibliography}
\end{document}